\documentclass{svproc}
\usepackage{url}
\usepackage{amsmath,amssymb,graphicx}
\usepackage{here}
\usepackage[dvipsnames]{xcolor}

\begin{document}

\title{String-inspired running-vacuum cosmology, quantum corrections and the current cosmological tensions}
\titlerunning{Quantum Stringy RVM and Tensions}

\author{Nick E. Mavromatos ({\it speaker})\inst{1,2} \and Joan Sol\`a Peracaula\inst{3} \and \\Adri\`a G\'omez-Valent\inst{4}}
\authorrunning{Nick E. Mavromatos et al.}
\tocauthor{Nick E. Mavromatos ({\it speaker}), Joan Sol\`a Peracaula, and Adri\`a G\'omez-Valent}

\institute{National Technical University of Athens, School of Applied Mathematical and Physical Sciences, Department of Physics, 9 Iroon Polytechniou Street, Zografou Campus GR157 80, Athens, Greece, \\
\email{mavroman@mail.ntua.gr}
\and
King's College London, Department of Physics, Strand, London WC2R 2LS, UK \\
\and
Departament de F\'\i sica Qu\`antica i Astrof\'\i sica,  and  Institute of Cosmos Sciences, Universitat de Barcelona, Av. Diagonal 647 E-08028 Barcelona, Catalonia, Spain.
\and
INFN, Sezione di Roma 2, and Dipartimento di Fisica, \\Universit\`a di Roma Tor Vergata,
via della Ricerca Scientifica 1, I-00133 Roma, Italy
}

\maketitle

\begin{abstract}
In the context of a string-inspired running vacuum model (RVM) of cosmology with anomalies and torsion-induced axion-like fields,
we discuss quantum corrections to the corresponding energy density, in approximately de Sitter eras, during which the Hubble parameter $H(t)$ varies very slowly with the  cosmic time $t$.
Such corrections arise either from graviton loops in the corresponding gravitational theory, or from path integration of massive quantum matter fields. They depend logarithmically on $H(t)$, in the form $H^n\, {\rm ln}(H^2)$, $n \in 2Z^+$. In the modern eras, for which the $n=2$ terns are dominant, such corrections may contribute to an alleviation of the currently observed cosmological $H_0$ and structure-growth tensions. In particular, we argue  that such an effect is accomplished for a (dynamically-broken) supergravity-based RVM cosmological model. In the current de-Sitter era, for this case, rather surprisingly, the quantum graviton corrections dominate those due to matter fields, provided the scale of the primordial (pre-RVM inflation) dynamical breaking of local supersymmetry lies near the reduced Planck scale, which is a natural assumption in the context of the model.
\end{abstract}


\section{Motivation and Summary}\label{sec:intro}

In this talk we shall review the work done in ref.~\cite{gms}, which deals with a phenomenological study of quantum corrections of string-inspired running vacuum models of cosmology.
Our main motivation  is an attempt to resolve, by modifying the standard $\Lambda$CDM Cosmology, the cosmological tensions in the current-epoch data~\cite{tensions,models}.  These are associated with either the measured value of the Hubble parameter today ($H_0$ tension),
using nearby galaxies and comparing with Cosmic Microwave Basckground (CMB) data 
using $\Lambda$CDM fits~\cite{Planck}, or growth-of-structure data.
Although there could be mundane astrophysical or statistical explanations  for these tensions~\cite{freedman}, nonetheless, their persisting nature triggered an enormous theoretical interest, as they might imply true, novel fundamental physics, deviations from the $\Lambda$CDM paradigm~\cite{models}.

One of the cosmological frameworks that offers potential, simultaneous, alleviations of both types of tensions is the running-vacuum-model (RVM) of cosmology, which was initially motivated  by semiqualitative arguments within the renormalization group ~\cite{rvm,rvm2,Solarvm2013,Sola:2015rra} and it has only recently been substantiated in full within the context of explicit quantum field theoretical  (QFT) calculations in curved spacetime\,\cite{qftrvm,qfteos,qftrvm2}. For a recent review of the modern formulation of the RVM, see \cite{Solarvm2022}.  According to the RVM, the entire evolution~\cite{lima}, but also thermodynamics (and entropy production)~\cite{rvmthermal}, of the Universe can be explained in terms of a smooth running of the cosmological vacuum energy $\rho_{\rm RVM}(t)$ with the cosmic time $t$. Due to {\it general covariance}, the vacuum energy density admits a perturbative expansion on {\it even} powers of the Hubble parameter $H$:\begin{align}\label{rvmener}
\rho_{\rm RVM} (H) = \frac{\Lambda (H^2)}{\kappa^2} = \frac{3}{\kappa^2} \Big( c_0 + \nu\, H^2 + \alpha\, \frac{H^4}{H_I^2} + \dots \Big),
\end{align}
where $\kappa^2 = 8\pi \rm G = M_{\rm Pl}^{-2}$ is the four-dimensional gravitational constant (with G the Newton's constant, and $M_{\rm Pl}=2.435 \times 10^{18}~{\rm GeV}$ the reduced Planck mass), $c_0, \nu, \alpha$ are real constants (and positive in the conventional RVM framework) dimensionless coefficients (with $H_I \sim 10^{-5} \, M_{\rm Pl}$ the inflationary scale, e.g. as measured by Planck Coll.~\cite{Planck}), and the $\dots$ denote higher even powers of $H$.\footnote{Actually there are also terms containing the cosmic time derivative of $H$, $\dot H$, but
such terms can be expressed in terms of the deceleration parameter $q$, as $\dot H = -(1 + q)\, H^2$,
and, since during each cosmological epoch $q$ is approximately constant, these terms can still be expressed in terms of $H^2$ so their effects can be absorbed (in each epoch) in  the coefficient of $H^2$ in \eqref{rvmener}. In detailed phenomenological analyses of RVM, the effects of $\dot H$ can be calculated explicitly~\cite{rvmpheno,rvmstruct,rvmtensions}. }
The constant $3c_0 \kappa^{-2}$ plays the r\^ole of a contribution to the cosmological constant, and the 
combination of today's values $3\kappa^{-2}(c_0  + \nu\, H_0^2 )$
is set to the current value of the observed dark energy density~\cite{Planck}, $\sim 10^{-122} M_{\rm Pl}^4$.

It should be remarked that the entire history of the Universe can be described by the truncation of the expansion \eqref{rvmener} to $H^4$ terms~\cite{lima}, although it should be noted that quantum field theory computations in RVM spacetimes lead to $H^2$ and $H^6$ terms in an explicit way, but not $H^4$~\cite{qftrvm,qftrvm2}. The latter can be produced by condensates of primordial chiral-gravitational-wave tensor modes  in string-inspired RVM models, called from now on stringy RVM (StRVM). The low-energy
gravitational theory stemming from such string theories is characterised by torsion (the r\^ole of which is played by the field strength of the antisymmetric tensor field in the massless gravitational string multiplet) and Chern-Simons (CS) gravitational anomalies~\cite{jackiw}, as discussed in detail in \cite{bms,ms}. The StRVM cosmological evolution is characterised by different phases, in which the coefficient $\nu$ in \eqref{rvmener} is negative during the inflationary era, and positive afterwards. Moreover, in the StRVM $c_0 =0$, since a cosmological constant seems not compatible with either  perturbative (scattering S-matrix)~\cite{smatrix} or non-perturbative string theory (swampland criteria)~\cite{swamp}. In our StRVM, approximate cosmological constant terms may arise either during the early RVM inflationary phase or at late eras, as a result of the formation of appropriate condensates~\cite{bms,ms}, which are {\it metastable} though,  and
eventually disappear. Such a metastability feature of the de Sitter phase seems to be also corroborated by local quantum field theory computations under appropriate vacuum renormalization in an expanding universe within a conventional RVM framework~\cite{qftrvm,qftrvm2}.
The RVM vacuum fluid is characterised by a de-Sitter type equation of state (EoS)~\cite{rvm,rvm2,Solarvm2013,Sola:2015rra,lima}:
$p_{\rm RVM} = - \rho_{\rm RVM}$, which is essentially valid at early Universe epochs, when the vacuum gravitational degrees of freedom are dominant. This is the case of the stringy RVM at early epochs~\cite{ms}.\footnote{We also note for completeness that similar situations, with time-dependent vacuum energy satisfying a de Sitter equation of state, have appeared in the context of some non-critical string models with brane defects~\cite{emndefects}.} In radiation and matter-dominance eras, one should consider the contributions to the energy density of such excitations, $\rho_m$ ($m$=radiation, matter), on top of the above vacuum contributions, which thus lead to a total energy density of the form $\rho_{\rm total} = \rho_{\rm RVM} + \rho_m $. The equations of state of such total contributions coincide approximately with the equation of state of the respective dominant component in the corresponding era. Such a result has been confirmed by detailed quantum field theory computations in RVM spacetime backgrounds~\cite{qfteos}, under appropriate renormalization  of the vacuum energy~\cite{qftrvm,qftrvm2}.

At late eras, the RVM framework is argued~\cite{rvmpheno,tsiapi} to be phenomenologically consistent with the plethora of the cosmological data available today~\cite{Planck}, including cosmic structure formation~\cite{rvmstruct}, but also indicating observable deviations from the $\Lambda$CDM paradigm, due to the $\nu H^2$ term, dominant at the late stages of the cosmological evolution. Fitting the current-era data with the RVM evolution framework leads to the estimate  $0 < \nu = \mathcal O(10^{-3})$. Curiously enough, the same estimate for $\nu$ is obtained by requiring consistency of the RVM framework with the Big-Bang-Nucleosynthesis data~\cite{bbnrvm}, upon assuming that the observed~\cite{Planck} current-era dark energy is of RVM type.
On the other hand, the $H^4$ (and higher order) terms in \eqref{rvmener}, which are dominant during the early eras of the cosmic evolution, drive an RVM inflation without the need for fundamental inflaton fields~\cite{lima,Sola:2015rra}. Moreover, there are claims~\cite{rvmtensions} that a variant of RVM, called RVM type II, allowing for a mild phenomenological dependence of the gravitational constant on the cosmic time $t$,  $\kappa \to \kappa (t)$, can simultaneously alleviate the
$H_0$ and growth tensions observed in late-cosmology data~\cite{tensions}.
As we shall argue in this talk, the StRVM may also provide a simultaneous resolution of both types of tensions, provided quantum corrections are taken into account. Such corrections are a consequence of quantum-gravity- induced logarithmic corrections $H^2{\rm ln}(H^2)$ to the modern era vacuum energy~\cite{mavrophil,ms} and, under appropriate conditions, could dominate, even in modern eras, the corresponding ones induced by appropriately renormalised quantum matter fields in the spacetime background of an expanding universe~\cite{qftrvm,qftrvm2}. However, a more detailed future study is required in order to ascertain whether the two types of RVM contributions (QFT versus stringy or quantum-gravity motivated effects)  are comparable in magnitude and reinforce each other or point towards different directions.

The structure of the talk is the following: in the next section \ref{sec:stringyRVM}, we formulate the StRVM gravitational model and discuss how, upon condensation of primordial chiral GW modes,
one obtains inflation of RVM type, without the need for external inflaton fields. In section \ref{sec:now} we discuss how in the current era, the StRVM, upon inclusion of appropriate quantum corrections, could contribute to a possible alleviation of the current-era $H_0$ tension, as well as tensions associated with galactic growth parameters \cite{gms}.

\section{Stringy Running Vacuum Model of Cosmology}\label{sec:stringyRVM}

The string-inspired cosmology model (StRVM) of \cite{bms,ms} is based on the bosonic part of the massless gravitational multiplet of the (closed sector of the) underlying microscopic superstring theory, which constitutes also the ground state~\cite{string}. This part consists of the fields of graviton (spin-2 symmetric tensor), dilaton (scalar, spin-0 field) and  spin-1 antisymmetric tensor (or Kalb-Ramond (KR)) gauge field, $B_{\mu\nu}=-B_{\nu\mu}$. In (3+1)-dimensions, after string compactification, which we restrict ourselves for the purposes of this talk, the dual to the field strength of the KR field,
$\mathcal H_{\mu\nu\rho}=  \kappa^{-1}\,  \partial_{[\mu}\,B_{\nu\rho]}$ (where
the $[\dots]$ denotes total antisymmetrization of the respective indices), is an axion-like particle, the so-called string-model independent axion~\cite{kaloper,svrcek}. The latter is essentially a Lagrange multiplier field implementing in the path integral a Bianchi identity constraint satisfied  by $\mathcal H_{\mu\nu\rho}$ after its modification to include Green-Schwarz anomaly cancellation terms~\cite{gs}:
\begin{equation}\label{modbianchi2}
 \varepsilon_{abc}^{\;\;\;\;\;\mu}\, {\mathcal H}^{abc}_{\;\;\;\;\;\; ;\mu}
 -  \frac{\alpha^\prime}{32\, \kappa} \, \sqrt{-g}\, R_{\mu\nu\rho\sigma}\, \widetilde R^{\mu\nu\rho\sigma}   =0,
\end{equation}
where where $\alpha^\prime = M_s^{-2}$ is the Regge slope of the string, with $M_s$ the string-mass scale, which is a free parameter in string theory, with $\sqrt{\alpha^\prime} \ne \kappa$ in general.
 The semicolon ; denotes covariant derivative with respect to the standard
Christoffel connection, $R_{\mu\nu\rho\sigma}$ denotes the Riemann tensor with respect to that connection,\footnote{Our conventions and definitions used throughout this work are those of \cite{bms}, that is: signature of metric $(+, -,-,- )$, Riemann Curvature tensor
$R^\lambda_{\,\,\,\,\mu \nu \sigma} = \partial_\nu \, \Gamma^\lambda_{\,\,\mu\sigma} + \Gamma^\rho_{\,\, \mu\sigma} \, \Gamma^\lambda_{\,\, \rho\nu} - (\nu \leftrightarrow \sigma)$, Ricci tensor $R_{\mu\nu} = R^\lambda_{\,\,\,\,\mu \lambda \nu}$, and Ricci scalar $R = R_{\mu\nu}g^{\mu\nu}$.} with $\widetilde R_{\mu\nu\rho\sigma} = \frac{1}{2} \varepsilon_{\mu\nu\lambda\pi} R_{\,\,\,\,\,\,\,\rho\sigma}$ the dual Riemann tensor, and
$\varepsilon_{\mu\nu\rho\sigma} = \sqrt{-g}\,  \epsilon_{\mu\nu\rho\sigma}$, $\varepsilon^{\mu\nu\rho\sigma} =\frac{{\rm sgn}(g)}{\sqrt{-g}}\,  \epsilon^{\mu\nu\rho\sigma}$,
$\epsilon^{0123} = +1$, {\emph etc.},  the gravitationally covariant Levi-Civita tensor densities, totally antisymmetric in their indices.

 The reader should note that the second term in left-hand side of \eqref{modbianchi2} is proportional to the gravitational anomaly~\cite{Eguchi,alvwitt}, which is an exact one-loop result, due to circulation of chiral degrees of freedom in quantum loops.

Ignoring the dilaton, which self-consistently can be set to zero~\cite{bms}, and path-integrating out the $\mathcal H$ field in the string effective action~\cite{string}, which, to $\mathcal O(\alpha^\prime)$ (quadratic order in a derivative expansion) we restrict ourselves here, behaves as a non-propagating field, one obtains the following action~\cite{kaloper,svrcek}
\begin{align}\label{sea3}
S^{\rm eff}_B =\; \int d^{4}x\sqrt{-g}\Big[ -\frac{1}{2\kappa^{2}} R + \frac{1}{2}\, \partial_\mu b \, \partial^\mu b +  \sqrt{\frac{2}{3}}\frac{\alpha^\prime}{96\, \kappa} b(x) R_{\mu\nu\rho\sigma}\, \widetilde R^{\mu\nu\rho\sigma} + \dots \Big],
\end{align}
where the dots $\dots$ denote higher-derivative, terms appearing in the string effective action, that we ignore for our discussion here. The field $b(x)$ is the pseudoscalar Lagrange multiplier which implements the Bianchi constraint \eqref{modbianchi2} in the path integral of the string-inspired (low-energy) effective gravitational theory. It is called gravitational, or KR, or, in modern string-theory language, string-model independent axion~\cite{svrcek}, as it characterises all string theories, independent of their types of compactified geometries.
Because to $\mathcal O(\alpha^\prime)$, one can view $\kappa^{-1} \, \mathcal H_{\mu\nu\rho}$ as a totally antisymmetric contorsion tensor, in the sense that the quadratic $\mathcal H_{\mu\nu\rho}\, \mathcal H^{\mu\nu\rho}$ in the effective low-energy gravitational lagrangian stemming from strings~\cite{string} can be absorbed in a generalised curvature with torsion~\cite{torsion}, one may say that the KR axion in this model has a geometric origin~\cite{mavrophil}.
 In string theory, there are of course many more axions, arising from compactification, which may lead to a rich phenomenology~\cite{arvanitaki} and cosmology~\cite{marsh}.
The model \eqref{sea3} constitutes a CS modification of general relativity (GR)~\cite{jackiw}.
and in general appears in theories with torsion, such as Einstein-Cartan quantum electrodynamics~\cite{kaloper,mavrophil}, with the $b$ field playing the r\^ole of the dual of the totally antisymmetric part of the torsion, as in the string theory case.

In the cosmology model of \cite{bms,ms} it is further assumed that there was a very early phase in the primordial Universe evolution, after the Big Bang, during which the spin 3/2 supersymmetric partners of gravitons (which exist in the superstring case), the gravitinos, $\psi_\mu$, could condense to scalar condensates $\sigma=\langle \overline{\psi}_\mu \, \psi^\mu \rangle \ne 0$, thus acquiring masses close to the Planck scale, and therefore breaking local supersymmetry (supergravity) dynamically~\cite{houston}. In such a scenario, there could be a primordial first hill-top inflation~\cite{ellisinfl}, which is not necessarily slow roll, as it does not lead to observable consequences other than  homogeneity and isotropy, which can then be used to describe quantitatively the transition, via tunnelling~\cite{ms}, of the isotropic and homogeneous system of massless gravitons and KR axions to the RVM second inflationary phase~\cite{ms}, that we now proceed to discuss.

After the dynamical local supersymmetry breaking, and the exit from the first hill-top inflation, gravitational waves can form, as a consequence of a lifting of the degeneracy of the vacua of the double-well potential of the gravitino condensate due to
dynamical percolation effects in the early Universe, which result in asymmetric occupation numbers of the two vacua~\cite{ross}. Such a lifting may lead to the formation of domain walls, whose asymmetric collapse and/or collisions can lead to the formation of primordial {\it chiral} (left-right asymmetric)  gravitational waves (GW). The latter can then condense, leading in turn to non-trivial condensates of the CS gravitational anomaly terms.

If we assume $\mathcal N(t)$ sources of such chiral primordial GW~\cite{mavlorentz}, then it can be shown, following~\cite{lyth}, that the GW induced CS condensate can be estimated as (assuming initially an approximately constant $H$, {\it i.e.} inflation, which we shall argue later that it is a consistent solution of the pertinent cosmic evolution):
 \begin{align}\label{condensateN2}
\langle R_{\mu\nu\rho\sigma} \, \widetilde R^{\mu\nu\rho\sigma} \rangle
=\frac{\mathcal N(t)}{\sqrt{-g}}  \, \frac{1.1}{\pi^2} \,
\Big(\frac{H}{M_{\rm Pl}}\Big)^3 \, \mu^4\, \frac{\dot b(t)}{M_s^{2}} \equiv n_\star \, \frac{1.1}{\pi^2} \,
\Big(\frac{H}{M_{\rm Pl}}\Big)^3 \, \mu^4\, \frac{\dot b(t)}{M_s^{2}}~,
\end{align}
where the overdot denotes derivative with respect to the cosmic time, and $n_\star \equiv \mathcal N(t)/\sqrt{-g}$ is the proper density of sources of GW, which we may assume to be approximately constant during the RVM inflation for simplicity and concreteness. The quantity $\mu$ is the UV cutoff of the effective low-energy theory, which serves as an upper bound for the momenta of the graviton modes that are integrated over in the computation of the condensate \eqref{condensateN2} in  the presence of chiral GW. To arrive at \eqref{condensateN2}, we have assumed isotropy and homogeneity of the string Universe, which can be guaranteed by the first hill-top inflation in the pre-RVM-inflationary scenario of \cite{ms}, as discussed above.\footnote{It should be stressed that the
estimate \eqref{condensateN2} is a valid estimate only in the field theory low-energy limit of the corresponding string theory. Unfortunately, as the estimate relies on the dominant UV physics near the cutoff $\mu$, in the case of microscopic string theory models it is the entire towers of massive string states that contribute, which makes an accurate estimation of the CS anomaly condensate not possible at present, apart from ensuring its non vanishing value.}

The explicit computations of \cite{bms,ms,mavlorentz} have shown that the total vacuum energy density during inflation assumes a RVM form \eqref{rvmener}:
\begin{align}\label{totalenerden}
\rho^{\rm total}_{\rm vac} =  -\frac{1}{2}\, \epsilon \, M_{\rm Pl}^2\, H^2 + 4.3 \times 10^{10} \, \sqrt{\epsilon}\,
\frac{|\overline b(0)|}{M_{\rm Pl}} \, H^4\,,
\end{align}
where we have used the following parametrisation of the approximately constant $\dot b$ axion background during inflation~\cite{bms}:
\begin{align}\label{axionbackgr}
b(t)=\overline{b}(t_0) + \sqrt{2\epsilon} \, H \, (t-t_0) \, M_{\rm Pl}\,,  \quad \overline{b}(t_0) < 0\,,
\end{align}
where $0 < \epsilon < 1 $ is a phenomenological parameter, and the time $t_0$ corresponds to the onset of the RVM inflation, implying that $\overline b(t_0)$ plays the r\^ole of a boundary condition for the KR axion.  In arriving at the estimates of \eqref{totalenerden} we took into account the fact that the assumption on the approximate de Sitter (positive-cosmological-constant) nature of the condensate, so as to lead to an approximately constant $H \simeq H_I$ during the RVM inflation, requires~\cite{bms,ms}:
$|\overline{b}(t_0)| \ge N_e \, \sqrt{2\epsilon} \, M_{\rm Pl}\, = \mathcal O(10^2)\,\sqrt{\epsilon}\, M_{\rm Pl}$, with $N_e= \mathcal O(60-70)$ the number of e-foldings of the RVM inflation.
The EoS of the string-inspired model during this inflationary phase has been explicitly computed in
\cite{ms} and found to coincide with the RVM EoS~\cite{lima,Sola:2015rra}, which justifies {\it a posteriori} calling this model {\it Stringy RVM}.  Because of the RVM form of the vacuum energy density \eqref{totalenerden}, we observe that it is the fourth power $H^4$ of the Hubble parameter, being the dominant one in the early Universe, that drives inflation in this model, which thus is of RVM form, not requiring external inflaton fields for its realization. Notice that any even power of $H$ beyond $2$ would  trigger inflation through that mechanism.  However, in the works \cite{rvmthermal} the structure \eqref{rvmener} (or the generalized one with $H^{2n},  n > 2)$ was assumed phenomenologically only, until such form got substantiated  in the modern QFT formulation of the  RVM  presented in \cite{qftrvm,qftrvm2}, which encompasses the entire cosmic evolution. On the other hand, the explicit $H^4$ power in \eqref{totalenerden} is genuine in the stringy approach~\cite{bms,ms}, and can be accounted for in it. This justifies {\it a posteriori} our analysis on arriving at the RVM inflation from GW-induced anomaly condensates in the StRVM.

From \eqref{condensateN2} and the form of the action \eqref{sea3}, the reader should observe that the condensates breaks the shift symmetry of the axion $b(x)$, leading to a linear potential for this field, of the form encountered in string theory, due to completely different reasons~\cite{silver}. When string-world-sheet effects are taken into account, which lead to periodic modulations of the shift-symmetry-breaking potentials not only of the KR axion $b$, but also of the compactification axions~\cite{svrcek},
one can under appropriate circumstances, explained in \cite{stamou}, arrive at a situation in which
the density of primordial black holes (pBH) produced during the RVM inflation in the StRVM is significantly enhanced, This affects the spectrum of GW in the radiation era, but also leads to significant fractions of such pBH playing the r\^ole of components of dark matter~\cite{pBH}.

\section{Modern Eras: Quantum corrections in Stringy RVM and cosmological tensions}\label{sec:now}

A detailed scenario for the post inflationary eras of the stringy RVM has been described in \cite{bms,ms}, where we refer the interested readers. We only mention here that at the end of the RVM inflation, the decay of the metastable vacuum leads to the generation of chiral fermionic matter and radiation. The chiral matter generates its own gravitational CS terms, but also chiral global anomalies, associated with the gauge sector. The matter-generated gravitational anomalies  {\it cancel} the primordial ones due to the Green-Schwarz mechanism, but the chiral anomalies remain. It is possible then~\cite{bms,ms}, that during the cosmic epoch corresponding to energy scales that correspond to dominance of Quantum Chromodynamics (QCD) effects, instantons of the colour SU$_{\rm c}$(3) gauge group generate periodic potentials, and thus masses, for the KR (and also the other) axions in string models, which could thus play the r\^ole of components of dark matter.

We now remark that, in the context of our stringy RVM model~\cite{ms,mavrophil},
upon making explicit use of methods developed for the prototype primordial supergravity model~\cite{bmssugra,houston}, which characterises the early pre-RVM inflation epochs of StRVM,
and
integrating out graviton quantum fluctuations  in the path integral around approximately (late-epoch) de Sitter cosmological space-times, one obtains logarithmic one-loop corrections to the energy density
of the form
\begin{align}\label{qglncorr}
\delta \rho_{\rm RVM}^{\rm QG}  \propto d_1\, H^2\, {\rm ln}(H^2)\,,
\end{align}
where $H(t)$ is the mildly depending on the cosmic time $t$ Hubble parameter in the approximately de Sitter current epoch, and the constant coefficients $d_1$:
\begin{align}\label{cs}
 d_1 \propto \kappa^2 \mathcal E_0, \,\, {\rm or} \,\,\, d_1 \propto \kappa^2 \mathcal E_0 \, {\rm ln}(\kappa^4 |\mathcal E_0|)\,,
\end{align}
with $|\mathcal E_0|$ a bare (constant) vacuum energy density scale. In supergravity models $\mathcal E_0 < 0$, but in the absence of supersymmetry, as is the case of modern eras we are interested in, $\mathcal E_0$ could be positive. This will play an important r\^ole for the generic phenomenological analysis of the stringy RVM at late epochs~\cite{gms}.

Such logarithmic corrections to the vacuum energy density {\it also} appear in quantum field theories in Robertson-Walker space times, under appropriate renormalization group treatments~\cite{qftrvm,qftrvm2}.
Such matter-generated effects are expected to be present in the current era, since there is still appreciable matter content today ($\Omega_m \simeq 0.26$). It is therefore plausible that they have appreciable effects in
alleviating the cosmological tensions. Naively, one would expect that such matter effects dominate over the quantum gravity effects.
However, upon closer examination and comparison with \eqref{cs}, we see that this is highly dependent on the magnitude of $\mathcal E_0$. For concreteness let us consider the case of
scalar real matter fields of mass $m$, non-minimally coupled to gravity with a parameter $\xi \in  R$~\cite{qftrvm,qftrvm2},
with the conformal theory corresponding to $\xi=1/6$. In this case one obtains for the renormalised current era energy density in an expanding universe  spacetime with Hubble parameter $H(t)$:
\begin{align}\label{qftsc}
\rho_{\rm RVM}^{\rm vac~QFT}  (H) &= \rho_{\rm RVM}^0 + \frac{3\, \nu_{\rm eff} (H) }{\kappa^2} \, \Big(H^2 - H_0^2 \Big), \nonumber \\
\nu_{\rm eff} (H) &\simeq \frac{1}{16\pi^2} \Big(\xi - \frac{1}{6}\Big) \, (\kappa^2 \, m^2)\,  \, {\rm ln}\Big(\frac{m^2}{H^2}\Big)  \,
\end{align}
The quantity $\rho_{\rm RVM}	^0 $ is the standard current-era RVM energy density, which we associate with the measured value of the cosmological constant through the relation  $\Lambda=8\pi G \rho_{\rm RVM}	^0$~\cite{Solarvm2022}.
The result \eqref{qftsc}, is based on the assumption~\cite{qftrvm,qftrvm2} $|{\rm ln}(m^2/H^2)| \gg 1$. Qualitatively similar expressions characterise the fermion case~\cite{qftrvm2}. However, the reader should notice that the corrections \eqref{qftsc} differ from the quantum graviton logarithmic corrections to the energy density, in that the logarithms
appear in the combination
$\delta \rho_{\rm RVM}^{\rm QFT} \propto (H^2-H_0^2)\,(\kappa^2 \, m^2){\rm ln}\Big(\frac{m^2}{H^2}\Big)$,
which is small in the modern era for which $H^2 \to H_0^2$. Unlike their quantum-gravity counterparts, such corrections cannot be cast in the form $R\,{\rm ln}(\kappa^2 R)$, with $R \sim 12H^2$ the de Sitter Ricci scalar. Comparing $\delta \rho_{\rm RVM}^{\rm QFT}$ with \eqref{qglncorr}, \eqref{cs}, we observe that, depending on the magnitude of the bare cosmological constant $\mathcal E_0$, matter effects could be suppressed compared to their quantum-graviton counterparts, given the smallness of the factor $(\kappa^2 \, m^2)\, (H^2 - H_0^2) \ll (\kappa^2 \, m^2)\, H^2 $ at modern epochs for which $H \to H_0$. Thus, even in modern eras, a non negligible possibility exists that quantum gravity corrections to the energy density of the stringy RVM dominate over quantum-matter-induced effects.
Making the latter assumption, we remark that such corrections can be best parametrised by an effective gravitational model with effective action~\cite{mavrophil,gms}:
\begin{equation}\label{eq:action}
S=-\int d^4x\,\sqrt{-g}\,\left[c_0+ R\, \left(c_1+c_2\ln\left(\frac{R}{R_0}\right)\right)\right]+S_m\,.
\end{equation}
where $R_0=12\, H_0^2$ is the de Sitter curvature scalar today, and $S_m$ is a matter/radiation action. The parameter $c_1 = 1/(2\kappa^2) + \delta c_1$, with $\delta c_1$ describing quantum corrections to the Newton's constant~\cite{gms}. Unitarity of the gravity sector requires $c_1 > 0$ in our conventions
\begin{figure}[H]
\center
\includegraphics[width=26pc]{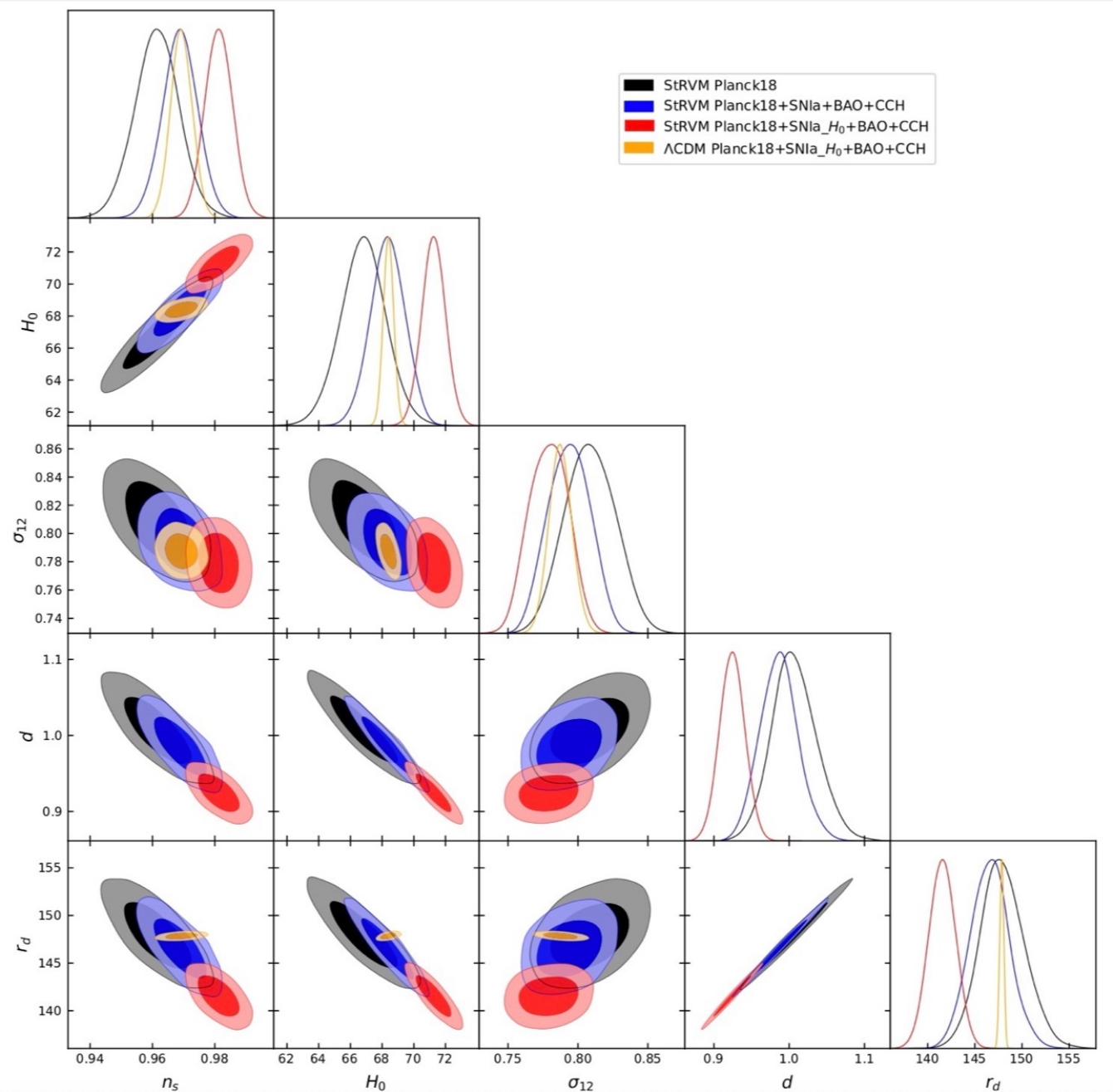}
\caption{The alleviation of the $H_0$ and growth tensions in the stringy RVM (StRVM) model of cosmology upon the inclusion of graviton quantum fluctuations at one-loop order in modern eras, whose incorporation makes the corresponding effective gravitational action taking the form \eqref{eq:action}. Here we quantify the amplitude of the matter power spectrum at linear scales with the parameter $\sigma_{12}=\sigma(r=12\,{\rm Mpc})$, i.e. the root-mean-square mass fluctuations at scales of $12$ Mpc. The reasons why the use of $\sigma_{12}$ rather than $\sigma_8=\sigma(r=8h^{-1} {\rm Mpc})$ is more appropriate for the StRVM are explained in \cite{gms}, which this figure is taken from. The $\Lambda$CDM predictions are depicted in orange colour.}
\label{fig:strvmH0}
\end{figure}
A full phenomenological analysis of this class of models, including fits to all available cosmological data at present (supernovae, CMB, Baryon Acoustic Oscillations and cosmic chronometers) is presented in ref.~\cite{gms}. An interesting case arises when one makes specific use of the broken-supergravity
phase of the StRVM.
On assuming that the supergravity contributions, corresponding to a primordial supersymmetry breaking scale $\sqrt{|f|}$, which contributes to the bare cosmological constant a {\it negative} value $-f^2 < 0$ (as required by supersymmetry~\cite{houston}), are the dominant ones of all the quantum-graviton-induced logarithmic corrections to the energy density from primordial to the current era,
so that the bare cosmological constant scale $\mathcal E_0$ appearing in \eqref{cs} coincides with $-\sqrt{f}< 0$,
we obtain~\cite{ms,mavrophil,gms}: 
\begin{align}\label{c12}
c_1 - c_2\, {\rm ln}(\kappa^2\, H_0^2) &=
\frac{1}{2\kappa^2} \Big[1 + \frac{1}{2}\, \kappa^4 \, f^2 \,\Big(0.083 - 0.049 \, {\rm ln}(3\kappa^4\, f^2)\Big)\Big]\, , \nonumber \\
c_2 &= - 0.0045\, \kappa^2 \, f^2 < 0\,.
\end{align}
We then constrain the constants $c_1,c_2$ from data~\cite{gms}, in order to alleviate the
$H_0$ and growth tensions, with the situation being summarised
in figure \ref{fig:strvmH0}, for the values of the dimensionless parameters $|\frac{c_2}{c_1 + c_2}| =
9 \times 10^{-3} \, \kappa^4 \, f^2 \lesssim \mathcal O(10^{-7})$ and $2\kappa^2\, (c_1 + c_2) = 0.924 \pm 0.017$.
This yields the following estimate on the magnitude of $\sqrt{|f|}$, $\sqrt{|f|} \kappa \sim 10^{-5/4} \lesssim 1$, implying a subplanckian scale for primordial dynamical supergravity breaking, consistent with the transplanckian conjecture.
As follows from \eqref{c12}, such values are also consistent with the perturbative modifications of $c_1$ from the (3+1)-dimensional gravitational constant $1/(2\kappa^2)$, despite the fact that $\kappa^2 H_0^2 = \mathcal O(10^{-122}) \ll 1$. With such scales one can also see~\cite{mavrophil,ms} that the quantum graviton corrections during the RVM inflationary eras are subleading as compared to the $H^4$ terms in the vacuum energy density \eqref{totalenerden} induced by GW condensates, and thus they do not affect our mechanism for inflation discussed in the previous section and in \cite{bms,ms}.

\section*{Acknowledgements} NEM would like to thank the organisers of
the 40th Conference on Recent Developments in High Energy Physics and Cosmology (HEP2023, 5-7 April 2023) of
the Hellenic High Energy Physics Society in the U. of Ioannina (Greece) for their kind invitation to give this plenary talk.
The work of NEM is supported in part by the UK Science and Technology Facilities research Council (STFC) under the research grants ST/T000759/1 and ST/X000753/1. The
work of JSP is funded in part by the projects PID2019-105614GB- C21, FPA2016-76005-C2-1-P (MINECO, Spain), 2021-SGR-249 (Generalitat de Catalunya) and CEX2019-000918-M (ICCUB, Barcelona). The work of AGV is funded by the Istituto Nazionale di Fisica Nucleare (INFN) through the project of the InDark INFN Special Initiative: “Dark Energy and Modified Gravity Models in the light of Low-Redshift Observations” (n. 22425/2020).
N.E.M. and J.S.P also acknowledge participation in the COST Association Action CA18108 ``{\it Quantum Gravity Phenomenology in the Multimessenger Approach (QG-MM)}''.
AGV and JSP take part in the COST Association Action CA21136 “Addressing observational tensions in cosmology with systematics and fundamental physics (CosmoVerse)”.


\begin{thebibliography}{99}

\bibitem{gms}
A.~G\'omez-Valent, N.~E.~Mavromatos and J.~Sol\`a Peracaula,
[arXiv:2305.15774 [gr-qc]].


\bibitem{tensions}
L.~Verde, T.~Treu and A.~G.~Riess,
Nature Astron. \textbf{3}, 891
[arXiv:1907.10625 [astro-ph.CO]].
L.~Perivolaropoulos and F.~Skara,
New Astron. Rev. \textbf{95} (2022), 101659.

\bibitem{models}
E.~Abdalla, \textit{et al.}
JHEAp \textbf{34} (2022), 49-211;
E.~Di Valentino, A.~Melchiorri, O.~Mena and S.~Vagnozzi,
Phys. Rev. D \textbf{101} (2020) no.6, 063502.



\bibitem{Planck}
N.~Aghanim \textit{et al.} [Planck],
Astron. Astrophys. \textbf{641} (2020), A6
[erratum: Astron. Astrophys. \textbf{652} (2021), C4].

\bibitem{freedman}
W.~L.~Freedman,
Nature Astron. \textbf{1} (2017), 0121.


\bibitem{rvm}
I.~L.~Shapiro and J.~Sol\`a,
JHEP \textbf{02} (2002), 006;
Phys. Lett. B \textbf{682} (2009), 105-113.

\bibitem{rvm2}
J.~Sol\`a, 
J.Phys.A {\bf 41} (2008) 164066.

\bibitem{Solarvm2013}
J.~Sol\`a,
J.Phys.Conf.Ser. {\bf 453} (2013) 012015;
AIP Conf.Proc. 1606 (2015) 1, 19-37.



\bibitem{Sola:2015rra}
J.~Sol\`a and A. G\' omez-Valent, Int.J.Mod.Phys.D{\bf 24} (2015) 1541003. 

\bibitem{qftrvm}
C.~Moreno-Pulido and J.~Sol\`a Peracaula,
Eur.Phys.J.C  \textbf{82} (2022) no.6, 551.
Eur.Phys.J.C  \textbf{80} (2020) no.8, 692.

\bibitem{qfteos}
C.~Moreno-Pulido and J.~Sol\`a Peracaula,
Eur. Phys. J. C \textbf{82} (2022) no.12, 1137.



\bibitem{qftrvm2} 
C.~Moreno-Pulido, J.~Sol\`a Peracaula and S.~Cheraghchi,
Eur.Phys.J.C {\bf 83} (2023) 637. 




\bibitem{Solarvm2022}
J.~Sol\`a Peracaula,
Phil. Trans. Roy. Soc. Lond. A \textbf{380} (2022), 20210182. 

\bibitem{lima}
E.~L.~D.~Perico, J.~A.~S.~Lima, S.~Basilakos and J.~Sol\`a,
Phys. Rev. D \textbf{88} (2013), 063531;
J.~A.~S.~Lima, S.~Basilakos and J.~Sol\`a,
Mon. Not. Roy. Astron. Soc. \textbf{431} (2013), 923-929.

\bibitem{rvmthermal} 
J.~A.~S.~Lima, S.~Basilakos and J.~Sol\`a,
Eur. Phys. J. C \textbf{76} (2016) no.4, 228;
Gen. Rel. Grav. \textbf{47} (2015), 40;
J.~Sol\`a Peracaula and H.~Yu,
Gen. Rel. Grav. \textbf{52} (2020) no.2, 17.

\bibitem{rvmpheno}
A.~G\'omez-Valent, J.~Sol\`a   and S. Basilakos,
JCAP {\bf 01} (2015) 004. 

\bibitem{rvmstruct}
A.~G\'omez-Valent and J.~Sol\`a,
Mon. Not. Roy. Astron. Soc. \textbf{448} (2015), 2810-2821;
A.~G\'omez-Valent, E.~Karimkhani and J.~Sol\`a,
JCAP \textbf{12} (2015), 048.

\bibitem{rvmtensions}
J.~Sol\`a Peracaula, A.~G\'omez-Valent, J.~de Cruz Perez and C.~Moreno-Pulido,
EPL \textbf{134} (2021) no.1, 19001;
 Universe {\bf 9} (2023) 6, 262. 
 
\bibitem{jackiw}
R.~Jackiw and S.~Y.~Pi,
Phys. Rev. D \textbf{68} (2003), 104012;
S.~Alexander and N.~Yunes,
Phys. Rept. \textbf{480} (2009), 1-55.



\bibitem{bms}
S.~Basilakos, N.~E.~Mavromatos and J.~Sol\`a Peracaula,
Phys. Rev. D \textbf{101} (2020) no.4, 045001;
Phys. Lett. B \textbf{803} (2020), 135342.

\bibitem{ms}
N.~E.~Mavromatos and J.~Sol\`a Peracaula,
Eur. Phys. J. Plus \textbf{136} (2021) no.11, 1152;
Eur. Phys. J. ST \textbf{230} (2021) no.9, 2077-2110.



\bibitem{smatrix}
S.~Hellerman, N.~Kaloper and L.~Susskind,
JHEP \textbf{06} (2001), 003;
W.~Fischler, A.~Kashani-Poor, R.~McNees and S.~Paban,
JHEP \textbf{07} (2001), 003.


\bibitem{swamp}
H.~Ooguri and C.~Vafa,
Nucl. Phys. B \textbf{766} (2007), 21-33;
G.~Obied, H.~Ooguri, L.~Spodyneiko and C.~Vafa,
[arXiv:1806.08362 [hep-th]];
S.~K.~Garg and C.~Krishnan,
JHEP \textbf{11} (2019), 075;
H.~Ooguri, E.~Palti, G.~Shiu and C.~Vafa,
Phys. Lett. B \textbf{788} (2019), 180-184;
For reviews see:
E.~Palti,
Fortsch. Phys. \textbf{67} (2019) no.6, 1900037,
and references therein.

\bibitem{emndefects}
J.~R.~Ellis, N.~E.~Mavromatos and D.~V.~Nanopoulos,
Gen. Rel. Grav. \textbf{32} (2000), 943-958.






\bibitem{tsiapi}
G.~Papagiannopoulos, P.~Tsiapi, S.~Basilakos and A.~Paliathanasis,
Eur. Phys. J. C \textbf{80} (2020) no.1, 55;
P.~Tsiapi and S.~Basilakos,
Mon. Not. Roy. Astron. Soc. \textbf{485} (2019) no.2, 2505-2510.





\bibitem{bbnrvm}
P.~Asimakis, S.~Basilakos, N.~E.~Mavromatos and E.~N.~Saridakis,
Phys. Rev. D \textbf{105} (2022) no.8, 8.

\bibitem{mavrophil}
N.~E.~Mavromatos,
Universe \textbf{7} (2021) no.12, 480;
Phil. Trans. A. Math. Phys. Eng. Sci. \textbf{380} (2022) no.2222, 20210188.


\bibitem{string}
M.~B.~Green, J.~H.~Schwarz and E.~Witten,
``Superstring Theory Vols. 1 \& 2: 25th Anniversary Edition,'' (Cambridge University Press, 2012)
ISBN 978-1-139-53477-2, 978-1-107-02911-8;
ISBN 978-1-139-53478-9, 978-1-107-02913-2.


\bibitem{kaloper}
M.~J.~Duncan, N.~Kaloper and K.~A.~Olive,
Nucl. Phys. B \textbf{387} (1992), 215-235.

\bibitem{svrcek}
P.~Svrcek and E.~Witten,
JHEP \textbf{06} (2006), 051.


\bibitem{gs}
M.~B.~Green and J.~H.~Schwarz,
Phys. Lett. B \textbf{149} (1984), 117-122.






\bibitem{Eguchi}
T.~Eguchi, P.~B.~Gilkey and A.~J.~Hanson,
Phys. Rept. \textbf{66} (1980), 213.

\bibitem{alvwitt} L.~Alvarez-Gaume and E.~Witten,
Nucl. Phys. B \textbf{234} (1984), 269.

\bibitem{torsion}
F.~W.~Hehl, P.~Von Der Heyde, G.~D.~Kerlick and J.~M.~Nester,
Rev. Mod. Phys. \textbf{48} (1976), 393-416;
I.~L.~Shapiro,
Phys. Rept. \textbf{357} (2002), 113.



\bibitem{arvanitaki}
A.~Arvanitaki, S.~Dimopoulos, S.~Dubovsky, N.~Kaloper and J.~March-Russell,
Phys. Rev. D \textbf{81} (2010), 123530.


\bibitem{marsh}
D.~J.~E.~Marsh,
Phys. Rept. \textbf{643} (2016), 1-79.

\bibitem{houston}
J.~Alexandre, N.~Houston and N.~E.~Mavromatos,
Phys. Rev. D \textbf{88} (2013), 125017;
Int. J. Mod. Phys. D \textbf{24} (2015) no.04, 1541004.

\bibitem{ellisinfl}
J.~Ellis and N.~E.~Mavromatos,
Phys. Rev. D \textbf{88} (2013) no.8, 085029.

\bibitem{ross}
Z.~Lalak, B.~A.~Ovrut and S.~Thomas,
Phys. Rev. D \textbf{51} (1995), 5456-5474;
Z.~Lalak, S.~Lola, B.~A.~Ovrut and G.~G.~Ross,
Nucl. Phys. B \textbf{434} (1995), 675-696;
D.~Coulson, Z.~Lalak and B.~A.~Ovrut,
Phys. Rev. D \textbf{53} (1996), 4237-4246.




\bibitem{mavlorentz}
N.~E.~Mavromatos,
``Lorentz Symmetry Violation in String-Inspired Effective Modified Gravity Theories,''
[arXiv:2205.07044 [hep-th]], contribution to {\it 740. WE-Heraeus-Seminar : Experimental Tests and Signatures of Modified and Quantum Gravity Workshop}, invited chapter to Springer book, in press.


\bibitem{lyth}  
S.~H.~S.~Alexander, M.~E.~Peskin and M.~M.~Sheikh-Jabbari,
Phys. Rev. Lett. \textbf{96} (2006), 081301;
D.~H.~Lyth, C.~Quimbay and Y.~Rodriguez,
JHEP \textbf{03} (2005), 016.

\bibitem{silver}
L.~McAllister, E.~Silverstein and A.~Westphal,
Phys. Rev. D \textbf{82} (2010), 046003.




\bibitem{stamou}
N.~E.~Mavromatos, V.~C.~Spanos and I.~D.~Stamou,
Phys. Rev. D \textbf{106} (2022) no.6, 063532.



\bibitem{pBH}
B.~Carr, F.~Kuhnel and M.~Sandstad,
Phys. Rev. D \textbf{94} (2016) no.8, 083504;
S.~Clesse and J.~Garc\'\i{}a-Bellido,
Phys. Dark Univ. \textbf{15} (2017), 142-147.

\bibitem{bmssugra}
S.~Basilakos, N.~E.~Mavromatos and J.~Sol\`a,
Universe \textbf{2} (2016) no.3, 14.

\end{thebibliography}
\end{document}